# Night sky quality monitoring in existing and planned dark sky parks by digital cameras

## Zoltán KOLLÁTH[1], Anita DÖMÉNY[1]

[1]Eötvös Loránd University, Savaria Department of Physics, Szombathely, 9700, Hungary

*Corresponding Author: Z. Kolláth (zkollath@gmail.com)

**Abstract**

A crucial part of the qualification of international dark sky places (IDSPs) is the objective measurement of night time sky luminance or radiance. Modern digital cameras provide an alternative way to perform all sky imaging either by a fisheye lens or by a mosaic image taken by a wide angle lens. Here we present a method for processing raw camera images to obtain calibrated measurements of sky quality. The comparison of the night sky quality of different European locations is also presented to demonstrate the use of our technique.

## 1. Introduction

The qualification of the night sky is not an easy task, especially at locations where the level of light pollution is low [1]. Sky luminance strongly depends on weather conditions and natural phenomena like airglow. A standard method for performing night sky brightness measurements is the use of Sky Quality Meters (SQMs – [2]). However, single measurements with this device provide only information about zenith luminance and possible light domes close to the horizon are hidden. A more precise measurement system for this task is the method introduced by the US National Park Service [3], where an astronomical CCD camera is used to produce a mosaic of the sky (see Figure 1.). Although this technique provides the most precise measurements, it is not available for all organizations or individuals who are interested in the formation of an IDSP. A cheaper and then a more feasible method for all sky photometry is the application of DSLR (Digital Single Lens Reflex) or MILC (Mirrorless Interchangeable-Lens Camera) digital cameras. The sensitivity and repeatability of recent cameras give the possibility of scientific quality measurements when thoroughly calibrated. Such a method was used in the application of the Hungarian dark sky parks in Zselic and Hortobágy [4].

We started sky quality monitoring with a digital camera in the Zselic Landscape Protection Area and in its neighborhood in 2008. Now we repeat the measurements with several cameras, included the one which was used during the first survey (the old camera was stored for years unused). Therefore, it is guaranteed that all the measurements during the 8 year long timespan are comparable. Our first results indicate that there were no significant changes in sky radiance inside the park.

The long term monitoring is especially important, since the public lighting in the vicinity of the dark sky park has undergone major changes during this period of time. For example, the high pressure sodium lighting was replaced with LED based luminaries in the largest city (Kaposvár) in the region. The ULR of the lighting has significantly decreased as reflected in the shape of the light dome. However, due to the **rebound effect (an economical phenomenon that improving energy efficiency may save less energy than expected)**, the effective light pollution is not decreased. Using the RGB information of the photography, the variation in correlated colour temperature is easily recognized in the light dome of the city. It indicates that scotopic measures of the light pollution in the dark sky park originated from Kaposvár are increased. Fortunately, it has no noticeable effects in the Zselic region. We plan further measurements in the Zselic region and the final conclusions will be published in a forthcoming paper.

The result of a similar lighting remodelling of another Hungarian city, Szekszárd was already published [5]. At that location the sky brightness is decreased after remodelling in all colour bands due to the improved shielding of the new luminaires. The improvement was significant in the red channel and in a less extent in the blue channel - as expected from the spectral change from Sodium Lighting to white LED's. These comparison measurements have also demonstrated the usefulness of digital camera imagery in light pollution research. Another application of our method was the quantification the nocturnal sky brightness at a freshwater lake in Germany [6], to help biomonitoring at the site.

We measured the brightness of the night sky with calibrated DSLR cameras with a fisheye lens at different dark sky locations in Europe. After presenting our method of evaluating digital camera images, we provide some examples of the use of the method on some selected locations.

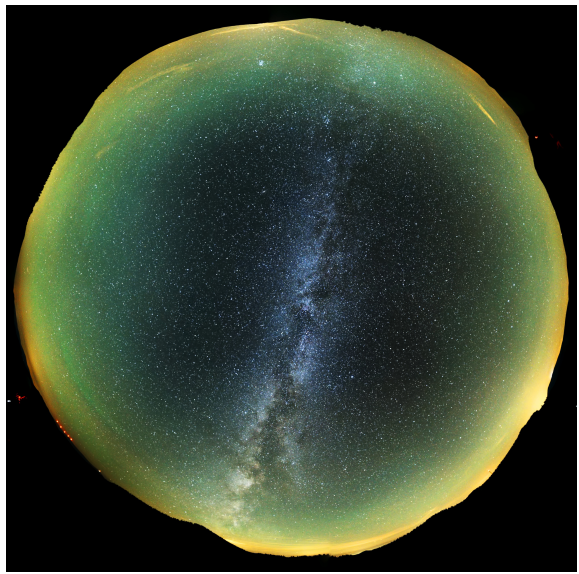

*Fig. 1. All sky mosaic composed of 10 individual exposures (Samyang 14mm f2.8 lens with Sony A7S camera). The strong green airglow is clearly visible. Location: Apuseni Nature Park, Padis.*

## 2. Methods

### 2.1. The software DiCaLum

Instead of developing a standalone software to analyze raw camera data, we decided to use some existing open source computer software to perform some major tasks of processing. The raw camera images are converted to standard PPM image files by dcraw [7]. The platform of our processing program is GNU Octave [8]. GNU Octave is a high-level language, developed for numerical computations. It has an extensive toolkit for manipulating and displaying matrices which are constructions that can represent images and radiance maps. Its computer language is mostly compatible with Matlab. GNU Octave is easily extensible via user-defined functions written in its own language. We used this property to develop DiCaLum (the name originates from the words: Digital Camera Luminance) a toolkit for night sky radiance and luminance measurements by DSLR and MILC cameras.

DiCaLum is a library of GNU Octave functions developed to read in digital camera RAW images and convert them to standard arrays containing the luminance or radiance values of the pixels. For example the following line:

```
[x , jd] = nsu("IMG_1111", "CR2", L_SIG45, C_60D);
```

reads in the raw image file 'IMG_1111.CR2'. The constants L_SIG45 and C_60D stand for the Sigma 4.5mm circular fisheye lens and for Canon EOS 60D camera respectively. These constants are defined for the lenses and cameras for which calibration data are available. The returned 'x' array contains the radiance values in natural sky units (NSU – see [10]). Using the matrix arithmetics of GNU Octave, it is easy to convert the NSU to other sky luminance/radiance units. For example the statement 'y = 21.6 - 2.5*log10(x)' converts the array to 'y' in mag/arcsec$^2$ units. Additional functions are available to get different kind of parameters, maxima, minima, mean values, etc. from the images. The simplest code to plot a false colour image is the following:

```
dicalum;
[x , jd] = nsu("IMG_1111", "CR2", L_SIG45, C_60D);
imagesc(log10(x),[-1,1]); colorbar;
```

This short code segment reads in the raw image file, then plots it in a logarithmic false color scale in the 0.1-10NSU range.

The plotting toolkit of Octave can be used to display the luminance map, or any derivative quantities like the dependence of luminance on zenith angle. We provide sample scripts to demonstrate the use of these functions. The user can develop her or his own scripts for additional tasks relatively easily by standard Octave commands.

## 2.2. Calibration database in DiCaLum

The functions provided in DiCaLum are useless without calibration data. Both the lens and the camera should be calibrated. In the program package we provide calibration values for the devices we used. A major factor is the vignetting effect of the lens: pixel values at the corners (or at the extremes of fisheye images) can be a factor of 4-10 lower compared to the central pixel values even if the luminance is uniform. [9] tested the vignetting effect of two identical lenses mounted on two identical cameras and they demonstrated that vignetting curves determined for one device can be reasonably used to correct the vignetting effect of the other device of the same brand.

There are different methods to fit the vignetting curve of a lens. A traditional method is to use an integrating sphere which provides uniform luminance in the whole field of view. However, this method depends on a high quality integrating sphere with a size that is compatible with the size of the lens. In the studies presented in [9] 49 white and gray squares were arranged in a semi-circle, and the luminance was measured in parallel with the photographic exposure. Then based on the measured luminance values of the squares the relative luminance of each square and so the vignetting curve can be calculated. The authors provided the 6th order polynomial fits of the vignetting measurements.

Instead of multiple patches of squares and a fixed camera, we used a single stabilized light source and rotated the camera to obtain the relative luminance values at different locations in the field of view. Our source is a set of six white LEDs with a double diffusor. This provides an almost homogeneous luminance on a 45 mm diameter disk. Our standard procedure is to take 200-300 images at a distance of 4-5 meters from the source in a dark room. The lens is focused to infinity, which provides a slightly out of focus images. Figure 2. displays a typical measurement sequence. Here the individual images were negated and plot on top of each other. For the processing we used DiCaLum itself: all images were processed separately, the location of the center of the light source and its mean instrumental brightness was determined. From the sensor coordinate and relative luminance the determination of the vignetting curve is a straightforward task.

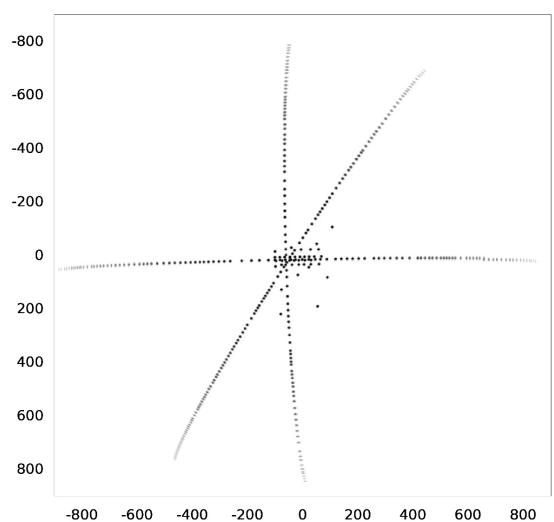

*Fig. 2. The negative images of the vignetting calibration source plotted on top of each other. The axes represent the pixel distance from the image center (Lensbaby 5.8mm fisheye lens attached to a SONY A7S camera).*

Although it is possible to fit the measured vignetting curve by a polynomial, the order of the polynomial needs to be high for a precise fit, at least for fisheye lenses. In the GNU Octave system, it is more feasible to use a spline fit to the data and construct a piecewise polynomial structure. We used a cubic spline with 6 knots – the locations of the inside knots were selected to provide optimal fit to the data.

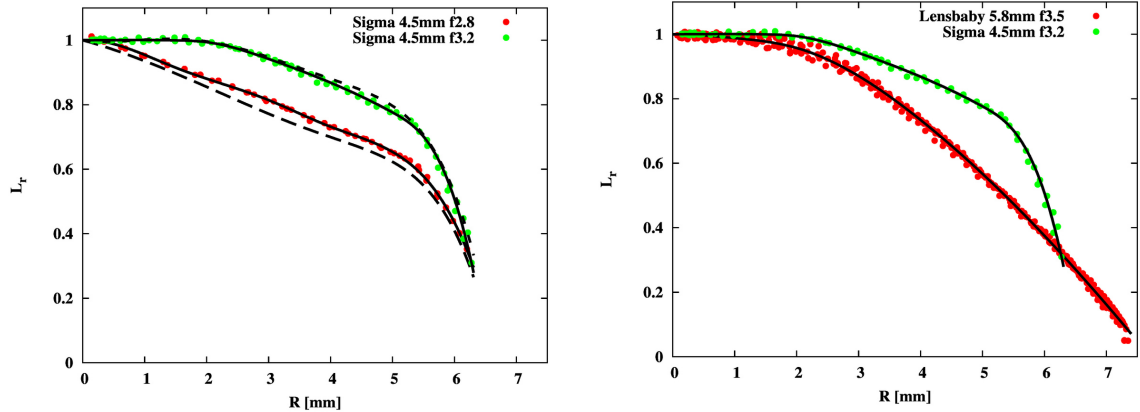

*Fig. 3. The relative luminance (Lr) as a function of the distance from the image center of the sensor. Left panel: Sigma 4.5mm lens with f2.8 and f3.2 aperture. Solid line: spline fit to our data, dashed line: polynomial fit presented in [9]. Right panel: Comparison of the vignetting curve of two circular fisheye lenses.*

Piecewise polynomial structures are handled in a straightforward way in GNU Octave, then it is easy to code the vignetting correction. Figure 3. displays the vignetting measurements and their fits for two circular fisheye lenses: the Sigma 4.5mm F2.8 EX DC HSM and the Lensbaby 5.8mm f/3.5. For the 4.5mm Sigma lens we provide data for the full open (f2.8) and the one stop closed (f3.2) apertures. The 5.8 mm Lensbaby lens is a manual one with now 'clicks' on the aperture ring, so it is recommended to use only the wide open setting – otherwise the repeatability of the measurement is not guaranteed. In general, the lenses have less vignetting when stopped down (higher f-number), but for the low light level measurements the widest aperture is needed to reduce exposure time. For comparison, we present the 6[th] order polynomial vignetting curve published in [9] by dashed curves in the left panel of Figure 3. At f3.2 the agreement of the vignetting curves is excellent - at f2.8 it reaches 5% at the center of the curve.

The calibration of the camera is performed with a lens already corrected for the vignetting effect. We prefer to use high quality luminance meters to calibrate the camera system in laboratory conditions. Then the camera is calibrated for a given set of light-sources in standard photometric (cd/m$^2$) units. The other possible way to calibrate the camera is performing astronomical photometry with a telephoto lens and processing this data to get the sky background in mag/arcsec$^2$ units. We prefer the calibration in laboratory and plan to do the calibration with a sensitive spectroradiometer and different type of sources. In this way the color dependence of the measurements can be estimated more precisely.

## 3. Test results

To demonstrate the usefulness of digital camera measurements, we selected two locations with recent observations. These locations are the Montsec Astronomical Park in Spain and Padis Area in the Apuseni Nature Park in Romania. Both locations have low light pollution levels and they have a similar elevation above the sea level. The New World Atlas of Artificial Sky Brightness [11] predict a little bit darker sky in Apuseni (dark blue color) than in Montsec (light blue color).

The luminance maps of the sky in false color are presented in Figure 4. The sky brightness as a function of zenith angle was also calculated. For the derivation of this function, we selected concentric rings at different zenith angles of a width of 2 degrees. For each rings the maximum, the minimum and the mean value of the radiance is calculated. On the right panels of Figure 4. we present all these three curves. This method provides a compact summary on the luminance distribution in the sky.

During the beginning of the first night at the Padis observations there was a very strong green airglow activity. It increased the sky luminance significantly everywhere on the sky. The upper two panels in Figure 4. clearly show the effect of increased airglow on sky luminance.

The zenith brightness at Montsec was excellent (around 22 mag/arcsec$^2$ with no Milky Way on the sky), but in the luminance map a clear gradient is visible. Close to the horizon, it is definitely brighter than at Padis. Around zenith the sky brightness is increased due to the presence of the Milky Way in the Padis measurements, although to some extend it can be handled with the minimum radiance values for all zenith angles, as they represent the portion of the sky which less affected by MW. The effect of the Milky Way is clearly displayed by the deviation of the maximum and minimum curves.

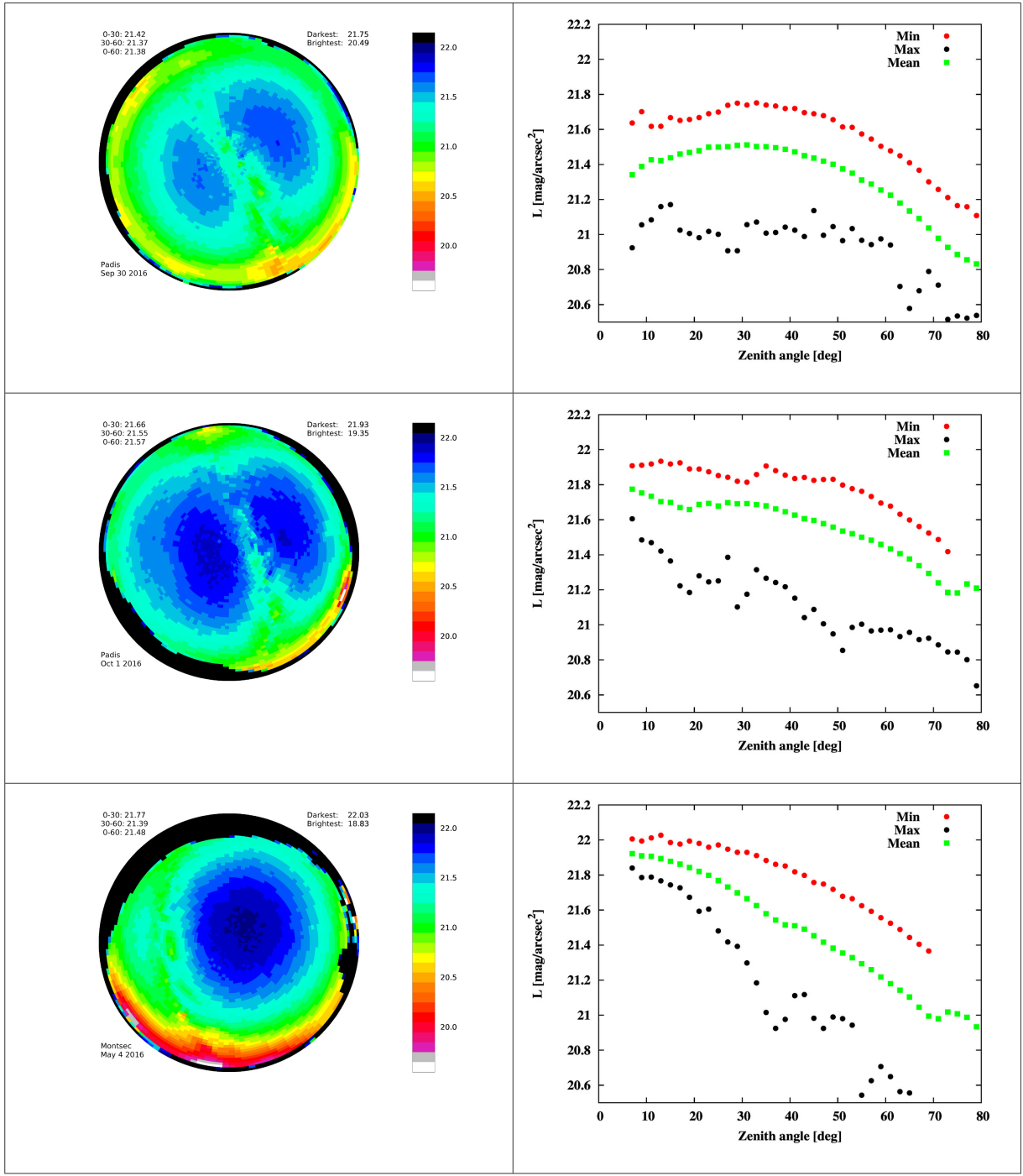

Fig. 4. Luminance maps calculated from the all-sky (right panels) and the variation of luminance vs zenith angle (right panels).

To compare the three cases presented in Figure 4, we also show the mean zenith angle - luminance curves in figure 5. The different gradient of sky radiance at the two locations is clearly demonstrated by comparing the Montsec and Padis curves. The difference can be originated from different weather and air conditions, although the available satellite data do not show any clear difference in the approximate aerosol optical depth for the two locations. Further radiative transfer modellings of the situations are also planned to understand the real mechanism behind the observed difference.

Figure 5. also shows the difference between the sky luminance values with different level of airglow activity. Even with low solar activity, we experienced a very strong effect on sky radiance. The variation of sky radiance due to the airglow is around 20% between zenith angles 30 and 70 degrees. The difference increases from 0.15 mag/arcsec² at 30 degrees to 0.25 mag/arcsec² at 70 degrees. The zenith values are also affected by the different location of the Milky Way, the airglow must have smaller effect there than the difference shown in the figure.

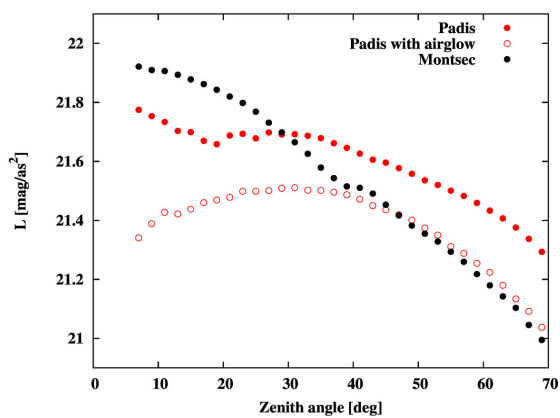

*Fig. 5. Mean sky brightness vs zenith angle – comparison of the three observations.*

**Conclusion**

We developed a computer tool (DiCaLum) , based on the GNU Octave programming language, to evaluate digital camera (DSLR or MILC) images. The basic functions of DiCaLum can be easily extended by standard GNU Octave programming. For example, such extensions provide publication quality plots of the observations. With some example measurements, it is demonstrated that:

- Digital cameras are capable of precise measurement of sky luminance/radiance; fisheye images or full sky mosaics with a calibrated system provide the full information about sky quality for possible dark sky parks.
- Effects of different geographic, weather conditions, or changes on the natural brightness (e.g. airglow) can be monitored with high precision by the method

We plan to distribute the code as a free software in the near future. The missing part is to make the full calibration of a wide range of cameras and lenses.

**Acknowledgements**

This work was partly supported by the EU COST Action ES1204 (Loss of the Night Network). The Montsec measurements are founded by the Stars4All awareness platform.